\documentclass[aps,pre,groupedaddress,showpacs,twocolumn,preprintnumbers,
amsmath,amssymb,floatfix]{revtex4}
\usepackage{graphicx}
\usepackage{bm}
\begin{document}
\author{Roya Zandi}

\affiliation{Department of Chemistry and Biochemistry, UCLA,
Box 951569, Los Angeles, California, 90095-1569}
\author{Joseph Rudnick}
\bibliographystyle{apsrev}

\affiliation{Department of Physics and Astronomy, UCLA, Box 951547,
Los Angeles, CA 90095-1547}

\author{Ramin Golestanian}

\affiliation{Institute for Advanced Studies in Basic Sciences, Zanjan
45195-159, Iran\\
Institute for Studies in Theoretical Physics and Mathematics, P.O.
Box 19395-5531, Tehran, Iran\\
Laboratoire de Physique de la Mati\`ere Condens\'ee, Coll\`ege de
France, UMR 7125 et FR 2438 du CNRS, 11 place Marcelin-Berthelot,
75231 Paris Cedex 05, France}

\date{\today}

\title{Anomalous Bending of a Polyelectrolyte}

\begin{abstract}
We report on a study of the shape of a stiff, charged rod that is
subjected to equal and opposite force couples at its two ends.  Unlike
a neutral elastic rod, which forms a constant curvature configuration
under such influences, the charged rod tends to flatten in the
interior and accumulate the curvature in the end points, to maximally
reduce the electrostatic self-repulsion.  The effect of this
nonuniform bending on the effective elasticity and on the statistical
conformations of a fluctuating charged rod is discussed.  An
alternative definition for the electrostatic persistence length is
suggested.  This new definition is found to be consistent with a
corresponding length that can be deduced from the end-to-end
distribution function of a fluctuating polyelectrolyte.

\end{abstract}


\maketitle

\section{Introduction and Summary} \label{sec:intro}

Given the ubiquity of charged linear structures in biology, and their
fundamental importance to essential processes in living systems, the
relationship of Coulomb interactions to the mechanical properties of
polyelectrolytes is a topic of pressing interest.  Indeed, there have
been a number of theoretical
\cite{joanny1,joanny2,hathiru,stevens,andelman,NOR} and
experimental \cite{eichinger,janmey,helfer} studies of the behavior of
charged polymeric chains, with an eye to elucidating the various
influences that control their equilibrium and statistical
characteristics.  In spite of the considerable effort expended, there
is, as yet, no comprehensive theoretical description of the way in
which charged chains respond to environmental influences.

The basic theoretical model of a polyelectrolyte chain (PE) is simplicity
itself.  A rod with an intrinsic elasticity quantified in terms of a
bending modulus carries charges, either uniformly distributed along
it, or concentrated at points along its axis.  The energy of this
chain consists entirely of the elastic energy associated with bending
of the rod and the electrostatic energy of interaction of the charges
on the rod.  The electrostatic interaction may be screened by
counterions in solution in the vicinity of the rod. This screening is
assumed to be Debye-like.


The characterization of the effects of intrinsic stiffness of a
neutral inextensible, or worm-like, chain (WLC) in terms of a
\emph{persistence length} is by now well-established \cite{landau}.
This quantity describes the exponential decay of correlations in the
orientation of the backbone of a WLC. It is directly related to the
energy stored in a short segment of the WLC which has been bent as the
result of the application of force couples at its end points.  An
extension of the persistence length to the case of a PE, due to Odijk
\cite{odijk} utilizes this basic approach to obtain that quantity in
the case of a stiff, charged rod.  In calculating the energy of the
bent segment of the PE, Odijk assumed that the shape of this segment
is not affected by the Coulomb interactions, an assumption shared by
all other expressions for the effective persistence length of a PE of
which we are currently aware \cite{fixman,khok,li}.  This leads to a
remarkably successful analytic expression for the persistence length
[see Eq.  (\ref{eqodijk}) in the next section].  This approximation
does not work in all regimes; in previous work, we have identified the
regimes in which this assumption is correct and other regimes in
which it fails to give rise to accurate answers \cite{rudnick,roya}.

The general question of the applicability of the notion of a
persistence length to a PE in the rod-like limit was explored in the
above mentioned work \cite{rudnick,roya} by the current authors.  This
study was conducted in the context of a calculation of the thermal
distribution of end-to-end distances of an ensemble of rod-like PE's.
It was found that there are regimes in which this distribution differs
substantially from the corresponding distribution for neutral WLC's.
When this is the case, the effective persistence length associated
with the end-to-end distance distribution (or radial distribution) is
not consistent with the formula derived in Ref.  \cite{odijk}.  The
effective persistence length is obtained in reference \cite{rudnick,roya} by
matching the PE distribution as closely as possible to the radial
distribution of uncharged worm-like chains \cite{frey}.

\begin{figure}
\includegraphics[height=1.5in]{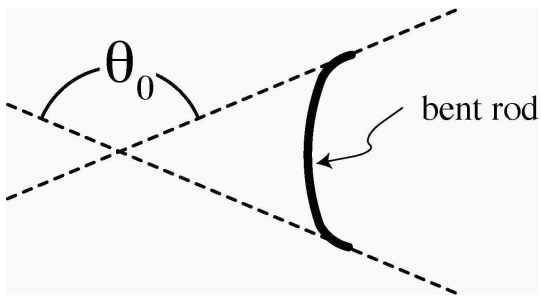}
\caption{The bent rod.  The angle $\theta_{0}$ between the two ends is
illustrated.  The relationship between the angle $\theta_{0}$ and the
energy, ${\cal E}$, stored the bent rod is displayed in Eq.
(\ref{lpdef}.}
\label{fig:fig0}
\end{figure}

In this paper, we undertake a calculation of the effective persistence
length.  Our approach is identical in overall philosophy to the that
utilized by Odijk \cite{odijk}.  A rod is subjected to external
torques that causes it to bend.  The energy, ${\cal E}$, in the bent
rod is calculated and related to, $ \theta_{0}$, the difference in
angles at the two ends, as shown in Fig.  \ref{fig:fig0}.
The precise relationship between the energy, ${\cal E}$, in the rod
and the angle $\theta_{0}$ is
\begin{equation}
\frac{\cal E}{k_{\rm B} T}=
\frac{\theta_{0}^{2}}{2}\frac{\ell_{p}}{L}. \label{lpdef}
\end{equation}

Here, $L$ is the total length of the rod.  In Odijk's approach, it is
assumed that the bent rod takes the form of an arc of a circle.  This
is true if all of the energy is elastic.  However, in the case of
interest here, a significant portion of the energy may be
electrostatic in nature.  The fundamental improvement over Odijk's
calculational method is that we determine the actual shape of the bent
rod, taking into account the effects of screened and unscreened
electrostatic interactions.  We assume that the shape taken is
controlled by the requirement that, in the absence of thermal
fluctuations, the rod minimizes its total energy.  The expression
obtained for the rod's shape is directly related to the inverse of the
Hamiltonian that relates the energy of the bent rod to its distortion
from a straight line.  This approach leads to new results for the
shape of the rod and to alterations in the energy and the persistence
length as defined in Eq.  (\ref{lpdef}).

\begin{figure}[tbp]
\includegraphics[height=1.5in]{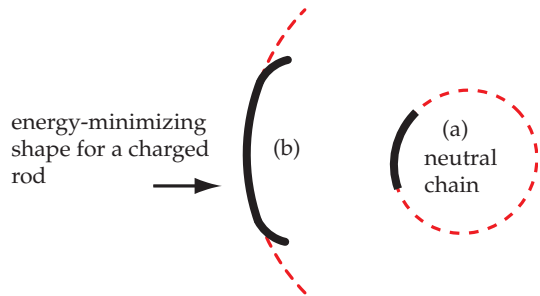} \caption{The energy-minimizing
shape of a charged rod with full treatment of electrostatic
interactions and end effects (plot (a)).  Plot (b) corresponds to the
energy-minimizing shape of a WLC.  A segment of a neutral WLC forms an arc
of a circle while the curvature of a PE segment can be concentrated at
its ends.}
\label{fig:fig1}
\end{figure}

We find that there are, in certain regimes, dramatic differences
between the shape of a short PE under the influence of force couples
at its ends and the shape of a similarly torqued WLC. In such regimes
the curvature of a PE is concentrated at its ends, while the WLC
distorts into a circular arc, as illustrated in Fig.  \ref{fig:fig1}.
In regimes in which the shapes of the PE and WLC coincide, we find
that the Odijk's formula for the effective persistence length is
accurate.  However, when the shape of the bent PE is inconsistent with
the assumption of constant curvature, the electrostatic persistence
length derived from Eq.  (\ref{lpdef}) deviates from Odijk's result.
On the other hand, this persistence length is compatible with the
persistence length deduced from the statistics of the conformations of
a fluctuating PE described by the radial distribution function
\cite{rudnick,roya}.

The rest of the paper is organized as follows.  Section
\ref{sec:model} introduces the basic elastic model for PE's, followed
by a general description on how one obtains the equilibrium shape of a
PE in Sec.  \ref{sec:strategy}.  The results for the shape of a PE are
presented in Sec.  \ref{sec:bending}, and it is shown in Sec.
\ref{sec:persistence} how a persistence length can be deduced from the
PE profile.  The suggested electrostatic persistence length is then
compared with Odijk's formula and also the persistence length that one
can extract from the radial distribution of the PE in Sec.
\ref{sec:radial}, which is then followed by concluding remarks in Sec.
\ref{sec:conc}.  Some details of the calculations are relegated to
appendices.

\section{Model Elasticity for Polyelectrolytes} \label{sec:model}

As noted above, we treat the polyelectrolyte as an inextensible
charged rod.  The total energy of such a rod is the sum of the
intrinsic elasticity and the screened electrostatic interaction
energy.  We assume that the electrostatic interaction is screened by
counterions that adjust more or less instantaneously to changes in the
overall shape of the polyelectrolyte.  Because of the inextensibility
of the PE's under consideration \cite{odijk,frey}, we adopt Kratky and
Porod WLC model to describe the bending energy of the chain
\cite{wlc}.  In this model, polymers are represented by a space curve
${\bf r}(s)$ as a function of the arc length parameter $s$.  The total
energy of the chain is given by
\begin{equation}
\frac{\cal E}{k_{\rm B} T} =\frac{\ell_{p0}}{2} \int_{0}^{L} d s
\left( \frac{d {\bf t}(s)}{ds}\right)^{2} +
\frac{\beta}{2}\int_{0}^{L} d s d s^{\prime} \frac{e^{-\kappa
|{\bf r}(s) - {\bf r}(s^{\prime})|}}{|{\bf r}(s) - {\bf
r}(s^{\prime})|}, \label{energy}
\end{equation}
where ${\bf t}$ is the unit tangent vector.  The second term in
the Eq. (\ref{energy}) is the Debye-H\"{u}ckel potential, in which
screening is controlled by the Debye length $\kappa^{-1}$ that is
a measure of the ionic strength of the solvent.  The quantity
$\beta=\ell_{\rm B}/b^2$ is the strength of the electrostatic
interaction with $b$ the average separation between neighboring
charges and $\ell_{\rm B}=e^{2}/\epsilon k_{B}T$ the Bjerrum
length. The quantity $\epsilon$ is the dielectric constant of the
ion-free solvent.

The chain is assumed to be sufficiently stiff that excluded volume
does not play a role.  We will consider PE's whose length, $L$, is
either comparable with or long compared to the intrinsic persistence
length, $\ell_{p0}$.  In both cases, we restrict our consideration to
regions in which the combination of intrinsic stiffness and repulsive
strength of the Coulomb interaction keeps the chains in their rodlike
limit \cite{en1}.  We do not take into account the fluctuation in the
charges localized to the chain and in the counterion system that can
give rise to attractive interactions leading to chain collapse
\cite{golestan,tanni}.

Orienting the tangent vector at one end of the PE, ${\bf t}(0)$, so
that it points in the positive ${\bf z}$ direction, we
characterize ${\bf t}(s)$ by two angles of rotation:
${\theta_{x}}(s) = \arctan[t_{x}(s)/\sqrt{1-t_x^2(s)-t_y^2(s)}]$
in the $xz$ plane and ${\theta_{y}}(s) =
\arctan[t_{y}(s)/\sqrt{1-t_x^2(s)-t_y^2(s)}]$ in the $yz$ plane.
For polyelectrolytes in the rodlike limit, ${\theta_{x}}(s)$ and
${\theta_{y}}(s)$ are small. With the help of the relation ${\bf
r}(s) -{\bf r}(s^{\prime})=\int_{s}^{s^{\prime}}d u \;{\bf t}(u)$,
we are able to expand both the bending energy and the screened
Coulomb interaction about the rodlike configuration up to
quadratic order in ${\vec
\theta}(s)=(\theta_{x}(s),\theta_{y}(s))$ following the Ref.
\cite{landau,odijk}.  In this case, Eq. (\ref{energy}) can be
written:

\begin{eqnarray}
\frac{\cal E}{k_{\rm B} T}& =& \frac{\ell_{p0}}{2} \int_{0}^{L}d s
\left\{\left( \frac{d \theta_{x}(s)}{ds}\right)^{2}+\left( \frac{d
\theta_{y}(s)}{ds}\right)^{2} \right\} \nonumber \\
&&+\frac{\beta}{2}\int_{0}^{L} \int_{0}^{L} ds \ ds^{\prime}{\cal
L}\left(s, s^{\prime} \right) \nonumber \\
&&\times \ \left\{ \theta_{x}(s)
\theta_{x}(s^{\prime})+\theta_{y}(s) \theta_{y}(s^{\prime})
\right\},
        \label{en1}
\end{eqnarray}

where the electrostatic kernel is given by
\begin{eqnarray}
&&{\cal L}(s,s')=\int_{0}^{L} d s_1 d s_2 \frac{[1+\kappa
(s_2-s_1)]
{\rm e}^{-\kappa (s_2-s_1)}}{(s_2-s_1)^3}\nonumber \\
&&\times \left\{(s_2-s_1) \left[\Theta(s-s_1)-\Theta(s-s_2)\right]
\delta(s-s')\right.\nonumber \\
&&-\left.\left[\Theta(s-s_1)-\Theta(s-s_2)\right]
\left[\Theta(s'-s_1)-\Theta(s'-s_2)\right]\right\},\label{L(s,s')}
\nonumber \\
\label{en2}
\end{eqnarray}
with $\Theta(s)$ the Heaviside step function.

Equations (\ref{en1}) and (\ref{en2}) constitute the expression for
the energy utilized by Odijk in his calculation of the electrostatic
persistence length.  The essence of this calculation is to constrain
the difference between the orientation of the tangent vectors at each
end of the rod, $\theta(L) - \theta(0) \equiv \theta_{0}$ and then to
determine the energy, ${\cal E}$ of the bent rod with the use of the
formulas in Eqs.  (\ref{en1}) and (\ref{en2}).  The persistence
length, $\ell_{p}$, is then  given by Eq. (\ref{lpdef}).
In his calculation of the total
energy of a bent segment of PE, Odijk makes the approximation that the
segment is characterized by a constant curvature \cite{odijk}.  That
is, he assumes that the electrostatic interaction does not cause the
shape of the PE to differ from that of a WLC segment.  According to
this picture, $\theta_{x,y}(s)$ are linear functions of $s$.  One can
then obtain an explicit, analytical expression for the total energy of
a bent polyelectrolyte segment, which leads directly to the following
prediction for the persistence length of such a charged rod through
Eq.  (\ref{lpdef}):
\begin{equation}
\ell_{p} = \ell_{p0} + \ell_{\rm Odijk}, \label{lpsum}
\end{equation}
where
\begin{eqnarray}
\ell_{\rm Odijk}&=&\frac{\beta L^2}{12}\left[e^{-\kappa
L}\left(\frac{1}{\kappa L}+\frac{5}{(\kappa L)^{2}}+\frac{8}{(\kappa
L)^{3}}\right) \right.  \nonumber \\
&&\left.+\frac{3}{(\kappa L)^{2}}-\frac{8}{(\kappa L)^{3}}\right].
\label{eqodijk}
\end{eqnarray}

In the next section, we sketch a general method of calculating the
equilibrium shape of elastic charged rods under the influence of
applied bending forces. This leads to a prescription for the
calculation of persistence length based on the energy of the bent PE.

\section{The general strategy for finding the energy-minimizing shape
of a bent polyelectrolyte}  \label{sec:strategy}

As noted previously, the principal advance in our approach to the
bending energy of a PE is that we calculate the actual shape of the
bent rod.  Here, we discuss the general approach to this calculation.
The energy of the rod in Eq.  (\ref{en1}) can be expressed as the
expectation value of the energy operator,${\cal H}$, i.e.
\begin{equation}
\frac{\cal E}{k_{\rm B} T} = \frac{1}{2} \int_{0}^{L} \int_{0}^{L}
ds ds^{\prime} \ \theta(s) {\cal H}(s,s^{\prime})
\theta(s^{\prime}).
         \label{enop}
\end{equation}
where ${\cal H}$, the bilinear ``energy'' operator, is equal to

\begin{equation}
{\cal H} = - \ell_{p0} \frac{d^{2}}{ds^{2}} \delta(s-s^{\prime}) +
\beta {\cal L}(s,s^{\prime}).
        \label{op1}
\end{equation}

To obtain the response of a polyelectrolyte to force couples at its
two ends, we minimize the energy expression in Eq.  (\ref{enop}) with
respect to $\theta_{x,y}(s)$, subject to the following boundary
conditions:

\begin{eqnarray}
\theta_{x}(0) & = & 0, \nonumber \\
\theta_{y}(0) & = & 0, \nonumber \\
\theta_{x}(L) & = & \theta_{0x}, \nonumber \\
\theta_{y}(L) & = & \theta_{0y}, \label{boundary}
\end{eqnarray}
where the angles ${\theta_{0x}}$ and ${\theta_{0y}}$ are assumed to be
very small, in order to keep the chain in the rod-like limit.  The
resulting Euler-Lagrange equations are completely decoupled with
respect to both variables ${\theta_{x}}(s)$ and ${\theta_{y}}(s)$;
therefore, we focus on planar deformations which can be characterized
entirely in terms of a single angle ${\theta}(s)$.  In this case, the
boundary conditions are simply,
\begin{eqnarray}
\theta(0) & = & 0, \nonumber \\
\theta(L) & = & \theta_{0}. \label{bound2}
\end{eqnarray}
We enforce these boundary conditions via Lagrange multipliers by adding
the terms
\begin{equation}
-\int_{0}^{L}\left\{\lambda_{0} \delta(s) \theta(s) -
\lambda_{L}\delta(s-L) \theta(s) \right\} ds,
      \label{bound3}
\end{equation}
to the energy in Eq. (\ref{enop}). Then we seek the solution that
minimizes the energy ${\cal E}$ in terms of an eigenfunction
expansion of the form
\begin{equation}
\theta_{\rm min}(s) = \sum_{n=0}^{\infty}c_{n} \psi_{n}(s),
        \label{ex1}
\end{equation}\noindent where $\psi_{n}(s)$ are eigenfunctions of the
bilinear Euler-Lagrange ``energy'' operator ${\cal H}$ in Eq.
(\ref{op1}).

The solution to the minimization equation, which is now in terms of
the amplitudes $c_{n}$, is
\begin{equation}
c_{j} = \frac{\lambda_{0} \psi_{j}(0)+ \lambda_{L}
\psi_{j}(L)}{\epsilon_{j}},
        \label{csol}
\end{equation}
where $\epsilon_{j}$'s are the eigenvalues of Eq.  (\ref{op1}).  The
values of $\epsilon_{j}$'s depend on the three dimensionless
parameters, $\beta L$, $\kappa L$, and $\ell_{p0}/L$.  Rotational
invariance implies that there must be one eigenfunction, $\psi_{0}(s)=
1/\sqrt{L}$, with eigenvalue $\epsilon_{0}=0$.  This is due to the
fact that simply tilting the rod, which yields a constant value for
$\theta(s)$, does not change the energy.  All other eigenfunctions
possess positive eigenvalues.  Equation (\ref{csol}) shows that there
is a singularity at $\epsilon_{0}=0$.  We can remove the singularity
by adding a term of the form $\frac {g}{2}\theta(s)^{2}$ to the energy
in Eq.  (\ref{enop}).  The parameter $g$ is a small ``gap'' parameter
that will be set equal to zero at the end.  The quantity $g$ is thus
added to the denominator of Eq.  (\ref{csol}), and there is no longer
a singularity at $\epsilon_{0}=0$ unless $g=0$.

We adjust the Lagrange multipliers, $\lambda_{0}$ and
$\lambda_{L}$, using the boundary conditions of Eq.
(\ref{bound2}). The two equations that lead to the results for the
$\lambda$'s are
\begin{eqnarray}
\theta(0) & = & \frac{1}{\sqrt{L}} \frac{\lambda_{0}+
\lambda_{L}}{g} + \sum_{n=1}^{\infty}\psi_{n}(0) \frac{\lambda_{0}
\psi_{n}(0)+ \lambda_{L} \psi_{n}(L)}{\epsilon_{n} + g} \nonumber
\\ &
= & 0,\label{theta0} \\
\theta(L) & = & \frac{1}{\sqrt{L}} \frac{\lambda_{0}+
\lambda_{L}}{g} + \sum_{n=1}^{\infty}\psi_{n}(L) \frac{\lambda_{0}
\psi_{n}(0)+ \lambda_{L} \psi_{n}(L)}{\epsilon_{n} + g} \nonumber
\\ & = & \theta_{0}.
        \label{thetaL}
\end{eqnarray}
The above equations reveal that the only possible way in which we
can obtain finite values for the angles at 0 and $L$ is to have
$\lambda_{L} \rightarrow - \lambda_{0}$ as $g \rightarrow 0$. Let
us set $\lambda_{0} + \lambda_{L} = g A \sqrt{L}$, where $A$ is a
constant that is set by adjusting boundary conditions.  The
general solution for $\theta(s)$ in the limit $g=0$ is, then,
\begin{equation}
\theta(s) = A + \lambda_{0}\sum_{n=1}^{\infty}\psi_{n}(s)
\frac{\psi_{n}(0) - \psi_{n}(L)}{\epsilon_{n}}.
        \label{gentheta}
\end{equation}

The limiting result of equal and opposite $\lambda$'s makes sense
if we think of those Lagrange multipliers in terms of torques, or
force couples, applied at the two ends of the PE segment.  Given
such a picture, we know that unless the two torques are equal and
opposite, there will be an uncontrolled rotation of the segment.
It is, in fact, possible to set up a calculation of the shape of a
segment under the influence of such torques.  Energy minimization
yields equations for the angles at the ends of the segment that
are precisely as given by Eqs. (\ref{theta0}) and (\ref{thetaL}),
with the $\lambda$'s proportional to the torques at each end.

In the next section we calculate numerically the energy-minimizing
shape of a charged rod and compare it with an arc of circle.  This
will lead to greater insight into the influence of energetics on the
classical shape of a bent segment of PE. It will also allow us to test
the fundamental relevance of an energy calculation such as the one
described above to the persistence length of a PE, defined in terms of
its conformational statistics.

\section{Nonuniform Bending of Charged Elastic Rods}    \label{sec:bending}

The expression for $\theta$ given in Eq.  (\ref{gentheta}) can also be
written as

\begin{equation}
\theta(s) = A + \lambda_{0}(K(s,L)-K(s,0)),
        \label{gentheta1}
\end{equation}
where
\begin{equation}
     K(s,s^{\prime})=\sum_{n=1}^{\infty} \frac{\psi_{n}(s)
\psi_{n}(s^{\prime}) }{\epsilon_{n}}, \label{invker}
\end{equation}

It is readily demonstrated that the quantity $K(s,s^{\prime})$
defined in (\ref{invker}) is the inverse of the energy operator in
Eq. (\ref{op1}).  The operator $K(s,s^{\prime})$ has been
calculated with the use of a cosine function basis set in Refs.
\cite{rudnick,roya}. We utilize our previously-obtained results
for the inverse operator to numerically calculate the quantity
$\theta$ in Eq.  (\ref{gentheta}). An outline of the construction
of the energy in this basis set is contained in Appendix
\ref{Coulomb}.

\begin{figure}[tbp]
\includegraphics[height=1.8in]{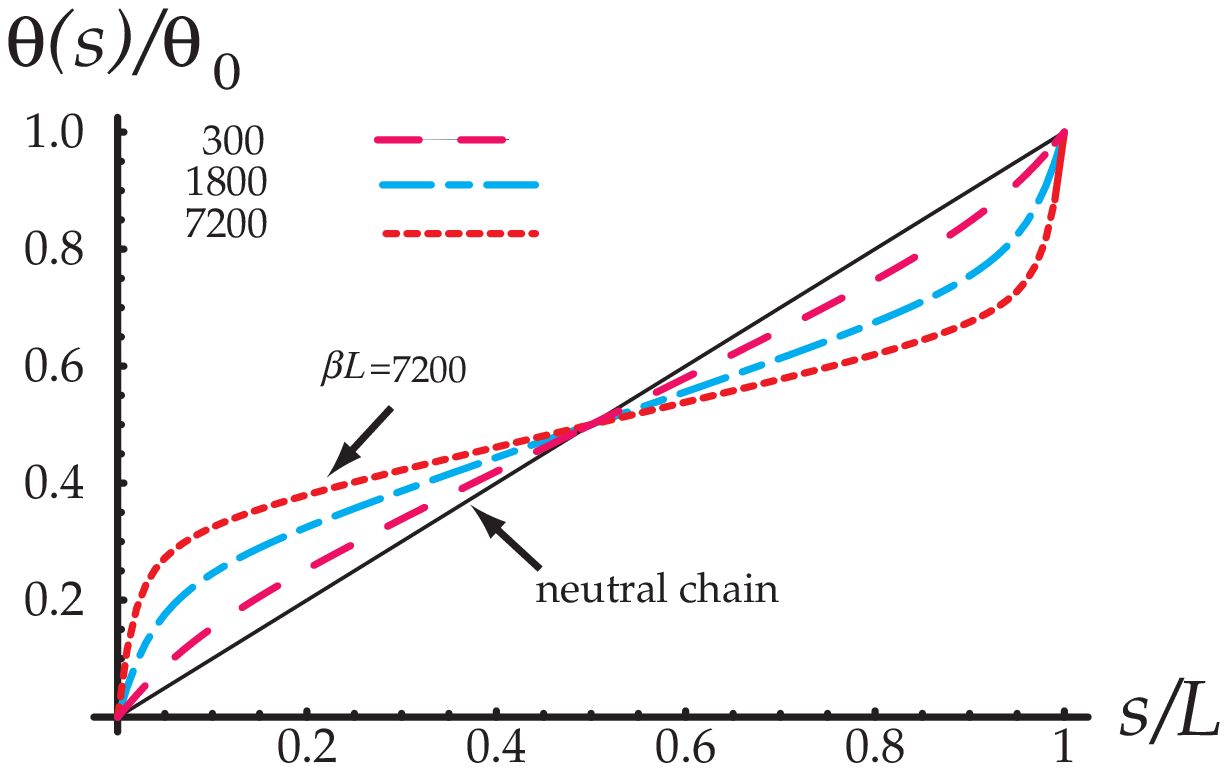} \caption{Plots of the
energy-minimizing shape for charged elastic rods [Eq.
(\ref{gentheta})] at $\kappa L=10$, $\ell_{p0}/L=0.5$ and $\beta
L=300$ (dashed line), $\beta L=1800$ (dashed and dotted line), $\beta
L=7200$ (dotted line).  The neutral chain forms an arc of a circle
(represented by a straight line).  As $\beta L$ increases, charged
chains tends to flatten more in the interior and accumulate curvature
in the end points. The extreme range of $\beta L$ is intended to
clearly indicate the influence of charging on shape.}
          \label{fig:theta1}
\end{figure}

In the absence of electrostatic interaction, a bent elastic rod
conforms to an arc of a circle, which is described by a linear
solution for Eq.  (\ref{gentheta1}).  In Fig.  \ref{fig:theta1}, the
energy-minimizing shapes of charged rods with $\kappa L=10$,
$\ell_{p0}/L=0.5$, and different values of $\beta L$, are compared
with that of a corresponding neutral chain.  As the figure clearly
illustrates, a $\theta(s)$ that depends linearly on arc length, $s$,
does not correspond to the minimum energy configuration of bent
charged elastic rods, which tend to flatten in the interior and
accumulate curvature in the exterior (see Fig.  \ref{fig:fig1}
above).  This is in part because the electrostatic self-repulsion is
lower in the end points due to the reduction in repelling neighboring
charges there.  Figure \ref{fig:theta1} also highlights the fact that
deviations from constant curvature become more pronounced as the
charging strength, $\beta L$ increase.  This tendency reinforces the
notion that Coulomb repulsion underlies the concentration of curvature
near the ends of the bent rod.

\begin{figure}[htb]
\includegraphics[height=1.8in]{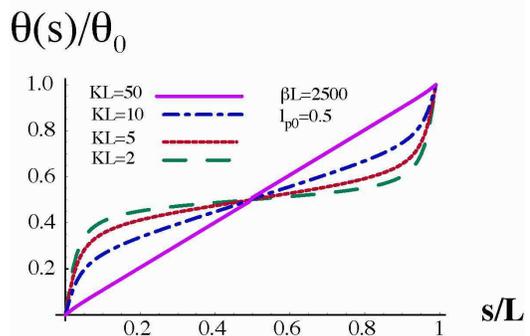}
\caption{Plot of $\theta(s)/\theta_{0}$ for a charged rod,
corresponding to various values of the inverse screening length,
$\kappa$. In all plots $\ell_{p0}=50$ nm, $\beta = 25$ nm$^{-1}$
(corresponding to DNA) and $L=100$ nm. The values of the screening
parameters are $\kappa=0.5$ nm$^{-1}$ (thick line), $\kappa=0.1$
nm$^{-1}$ (dashed and dotted line), $\kappa=0.05$ nm$^{-1}$
(dotted line), and $\kappa=0.02$ nm$^{-1}$ (dashed line),
respectively.} \label{fig:diffk}
\end{figure}

We have also investigated the influence of inverse screening
length, $\kappa$, and absolute length, $L$, on the shape of a bent
PE segment. Figure \ref{fig:diffk} illustrates the effects of a
change of $\kappa$ on the arclength dependence of $\theta$.  In
this figure, the bare persistence length $\ell_{p0}=50$ nm and the
charging parameter $\beta=25$ nm$^{-1}$ corresponding to DNA are
used, and the length of the segment is set equal to $100$ nm. It
is apparent that as the screening length increases the shape of
the bent segment deviates more and more from an arc of a circle.

\begin{figure}[htb]
\includegraphics[height=1.8in]{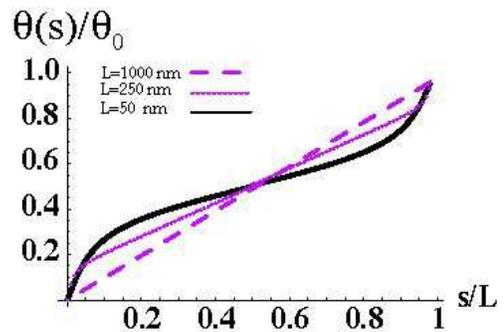}
\caption{Plot of $\theta(s)/\theta_{0}$ for a charged rod for
which the length $L$ is allowed to vary.  In all curves
$\ell_{p0}=50$ nm, $\beta = 25$ nm$^{-1}$ (corresponding to DNA)
and $\kappa=0.1$ nm$^{-1}$. It is evident that longer segments
behave more like a WLC in that when bent they take a shape with a
constant curvature. This reflects the influence on the shape of
the combination $\kappa L$, and is consistent with the tendencies
indicated in Fig. \ref{fig:diffk}.} \label{fig:difflength}
\end{figure}

Additionally, we have looked at the consequences on PE shape of
changes in the length of the segment, keeping all other parameters
fixed. Figure \ref{fig:difflength} shows how changing the length,
$L$, causes $\theta(s)$ to deviate from a straight line. As in
Fig. \ref{fig:diffk}, we have set $\ell_{p0}$ equal to 50 nm and
$\beta$ equal to 25 nm$^{-1}$, as corresponding to DNA. The
inverse screening length $\kappa$ in Fig. \ref{fig:difflength} is
fixed at 0.1 nm$^{-1}$.

\begin{figure}[tbp]
\includegraphics[height=2in]{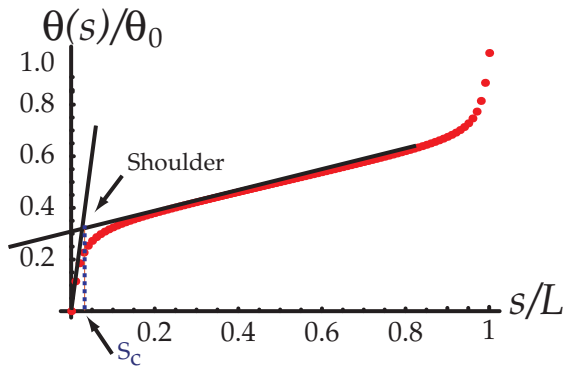}
\caption{The position of the shoulder $s_c$ can be obtained as the
crossover point of the tangents to the various segments of the
profile.} \label{fig:shoulder}
\end{figure}

The nonlocal electrostatic interaction seems to favor a decomposition
of the linear profile into a piece-wise linear one, in which the
interior takes up a lower curvature and the two end-segments in the
exterior acquire a higher curvature.  The relatively sharp changes in
the slope that result in ``shoulders'' in the profile take place
symmetrically at positions denoted by $s_c$ and $L-s_c$, which can be
defined in terms of the intersection points of the tangents to the
various segments of the profile, as illustrated in Fig.
\ref{fig:shoulder}.

\begin{figure}
\includegraphics[height=1.8in]{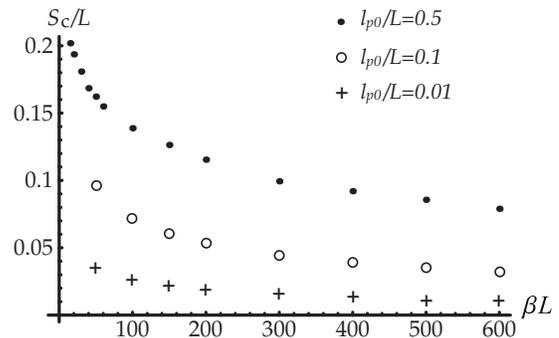} \caption{The position of
the shoulder $s_c$ as a function of the charging parameter $\beta L$,
for $\kappa L=0$ and $\ell_{p0}/L=$0.01 (crosses),0.1 (open circles)
and 0.5 (filled circles).  As expected, $s_c$ increases as the rod
becomes intrinsically more stiff.}
\label{fig:lpvalue}
\end{figure}

\begin{figure}[tbp]
\includegraphics[height=1.8in]{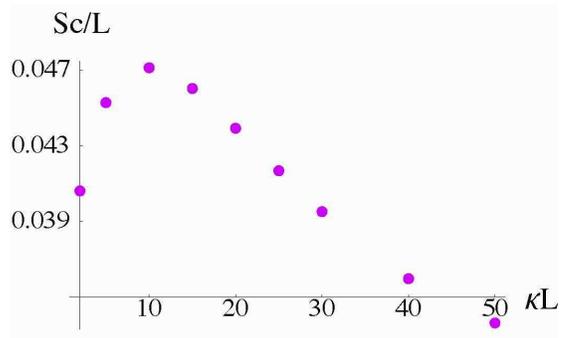} \caption{The position
of the shoulder $s_c$ as a function of the screening parameter
$\kappa L$, for $\ell_{p0}/L=0.5$ and $\beta L=2500$. The re-entrance
behavior is characterized by an initial slow increase followed by
a relatively faster decay at larger values of the screening parameter.}
\label{fig:newsc}
\end{figure}

The position of the shoulder $s_c$ is a monotonically decreasing
function of $\beta L$ as shown in Fig.  \ref{fig:lpvalue}.  The effect
of $\ell_{p0}/L$ on the $s_{c}$ is also illustrated in Fig.
\ref{fig:lpvalue}.  As seen in the figure, the shoulder $s_{c}$
increases upon increasing the intrinsic persistence length of the
chain.  The dependence of $s_c$ on screening, on the other hand,
appears to be more complicated.  One generally expects that as $\kappa
L$ increases, end-effects become less significant, and the value of
$s_{c}$ moves toward zero, resulting in a smoothing of the curvature
along the chain.  However, this is true only for strong screening.
Figure \ref{fig:newsc} represents the dependence of $s_{c}$ on $\kappa
L$ in the strong charging regime, where we observe that $s_c$ has a
relatively weak dependence on screening: $s_c$ slowly {\it increases}
as $\kappa L$ is increased and then starts decreasing with
further increase in $\kappa L$.

One can understand the appearance of the shoulder region as an end effect.
The nonlocal nature of the electrostatic self-interaction leads to enhanced
repulsion in the interior of the PE as compared to the end-segments.  This
is mainly due to the fact that there are fewer neighboring pairs at the
end-segments to contribute to the mutual repulsion.  In other words, one
might think of a crossover length scale at the two ends, below which the
intrinsic rigidity (that yields a local resistance to bending) dominates
the energetics of the chain, while beyond that length scale (i.e. in the
interior of the PE) it is the combination of the intrinsic rigidity and the
electrostatic repulsion that controls the energetics.  Interestingly, such
a crossover length scale has been introduced by Barrat and Joanny in their
study of the length-scale dependence of the PE rigidity
\cite{joanny1,joanny2}.  It is important to note that the Barrat-Joanny
crossover length is defined for the crossover in the fluctuations of the
angle $\langle\theta(s)^2\rangle$ (that also has a piece-wise linear
dependence on $s$).  Since the distribution of the angle $\theta(s)$ as
controlled by the Hamiltonian in Eq.  (\ref{enop}) is Gaussian, both
$\langle\theta(s)^2\rangle$ and the energy-minimizing $\theta$ (Eq.
\ref{gentheta1}) are linear functions of ${\cal H}(s,s^{\prime})^{-1}$ (Eq.
(\ref{op1})).  Thus we expect that in general the two crossover length
scales coincide.  Barrat and Joanny propose an expression for the crossover
length as

\begin{equation}
s_{c, \rm BJ}\simeq\sqrt{{\ell_{p0} \over \beta+4
\ell_{p0} \kappa^2}},  \label{BJSc}
\end{equation}
which exhibits the limiting forms of
$s_{c,\rm BJ}\sim\sqrt{\ell_{p0}/\beta}$ for
$\kappa \sqrt{\ell_{p0}/\beta} \ll 1$, and $s_{c, \rm BJ}\sim \kappa^{-1}$
for $\kappa \sqrt{\ell_{p0}/\beta} \gg 1$.

The Barrat-Joanny crossover length shows a qualitatively similar
behavior to the shoulder position $s_{c}$ as described above, except
for the slow initial increase in Fig.  \ref{fig:newsc} (for $s_{c}$ as
a function of $\kappa$).

There is a way, to reconcile this behavior with the Barrat-Joanny picture.
This can be achieved by considering the fact that in their derivation of
the expression in Eq.  (\ref{BJSc}) above, they have neglected a
logarithmic dependence in the electrostatic nonlocal kernel for technical
simplicity.  While it is not possible to calculate the correct crossover
length in a compact form as in Eq.  (\ref{BJSc}) when the logarithmic
factor is taken into account, one can extract the limiting forms of the
{\it augmented} Barrat-Joanny (aBJ) crossover length as 
\begin{equation}
s_{c,\rm aBJ}\sim\left\{ \begin{array} {cl} {1 \over \sqrt{\ln\left(\beta \over
\ell_{p0} \kappa^2 \right)}} \sqrt{\ell_{p0}/\beta} & \mbox{for $\kappa
\sqrt{\ell_{p0}/\beta} \ll 1$} , \\ \\ \kappa^{-1} & \mbox{for $\kappa
\sqrt{\ell_{p0}/\beta} \gg 1$}, \end{array} \right.  
\label{aBJSc}
\end{equation} which now exhibits an initial increase in qualitative
agreement with Figs.  \ref{fig:lpvalue} and \ref{fig:newsc}.  While
this picture can qualitatively account for the aforementioned
behaviors, we have not yet been able to achieve a quantitative
characterization of the shoulder position $s_{c}$ as a function of the
three dimensionless parameters $\ell_{p0}/L$, $\kappa L$, and $\beta
L$, and in particular compare it with the dependencies as suggested by
Eq.  (\ref{aBJSc}) above, due to the insufficiency of the numerical
data.


Having obtained the shape of a charged rod numerically, we can now
follow Odijk and calculate the persistence length of PE's.

\section{Derivation of the persistence length using Odijk's
method}\label{sec:persistence}

There is a straightforward way to calculate the energy of a bent rod,
based on the expression for the angle as a function of arc length.
The quantity $\theta(s)$ given in Eq.  (\ref{gentheta1}) is, to within
an additive constant, proportional to $K(s,L)-K(s,0)$, and the energy
of a bent charged rod can be written as
\begin{equation}
\frac{\cal E}{k_{\rm B} T}=\frac{\theta_{0}^{2}}{2} \
\frac{1}{K(0,0)-K(0,L)}.
         \label{eq:en3}
\end{equation}

\begin{figure}[tbp]
\includegraphics[height=1.7in]{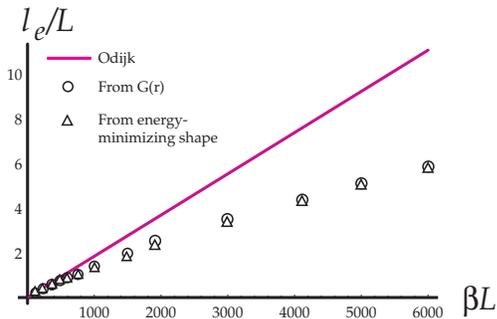}
\caption{The electrostatic persistence length,
$\ell_{e}$ obtained by Eq. (\ref{eq:l01}) for $\kappa L=10$ and
$\ell_{p0}/L=0.5$ at different values of $\beta L$, and comparison
with Odijk's persistence length as given by Eq. (\ref{eqodijk}),
and the electrostatic persistence length that can be deduced from
the radial distribution function of PE's \cite{rudnick,roya}. }
\label{fig:newper}
\end{figure}

Details of the calculation leading to Eq.  (\ref{eq:en3}) are
presented in the Appendix \ref{minenergy}.  According to the
definition of the persistence length in terms of the energy of the
bent rod, Eq.  (\ref{lpdef}), we have
\begin{equation}
\ell_{e} = \frac{L}{K(0,0)-K(0,L)}-\ell_{p0},
         \label{eq:l01}
\end{equation}
where $\ell_{e}$ is the electrostatic persistence length of the chain
as defined in the Sec.  \ref{sec:model}.  It is important to note that
the kernel $K(s,s^{\prime})$ depends on the parameters $\beta L$,
$\kappa L$, and $\ell_{p0}/L$ through the eigenvalues and
eigenfunctions of Eq.  (\ref{op1}).  Figure \ref{fig:newper} shows the
values of the electrostatic persistence length, $\ell_{e}$ obtained by
Eq.  (\ref{eq:l01}) for $\kappa L=10$ and $\ell_{p0}/L=0.5$ at
different values of $\beta L$ (triangles).  Odijk's persistence length
as given by Eq.  (\ref{eqodijk}) is also plotted in the figure for
comparison (solid line).  For small values of $\beta L$, $\ell_{e}$
coincides with the Odijk persistence length, $\ell_{\rm Odijk}$.  As
$\beta L$ increases, the deviation of $\ell_{e}$ and $\ell_{\rm
Odijk}$ becomes more significant.  The figure also displays the value
of the electrostatic persistence length that one can infer from the
distribution of end-to-end distances of an ensemble of fluctuating
rod-like PE segments (open circles) \cite{rudnick,roya}.  There will
be more on this subject in the next section.

Our general observation from the comparison of $\ell_{e}$ with
$\ell_{\rm Odijk}$ is that when the two quantities are equal, the bent
charged rod is close in shape to an arc of a circle.  That is,
$\theta(s)$ as a function of $s$ is nearly a straight line as in the
case of a neutral chain.  This indicates that as charging increases,
end-effects become more important and the description of PE's as
neutral chains with an adjusted persistence length is inappropriate.
It is clear that end-effects play a key role in the elasticity of
PE's.  Such effects are also apparent in the statistical conformations
of the charged rods.

\section{The influence of ``end-effects'' on the statistical conformation
of PE's}   \label{sec:radial}

The study of the end-to-end radial distribution function, $G({\bf
r})$, of a rod-like PE provides an excellent gauge of the
statistical conformation of polymers. Using the expression for
energy given in Eq. (\ref{energy}), we have obtained values for
the quantity
\begin{equation}
G({\bf r})=\langle\delta({\bf r}- {\bf R})\rangle,
                   \label{dist}
\end{equation}
where ${\bf R}={\bf r}(L)-{\bf r}(0)$. The average
in Eq. (\ref{dist}) is over an ensemble of PE chains. The function
$G({\bf r})$ is, then the probability that a given chain in the
ensemble will have an end-to-end distance equal to ${\bf r}$
\cite{rudnick,roya}.

\begin{figure}[tbp]
\includegraphics[height=2in]{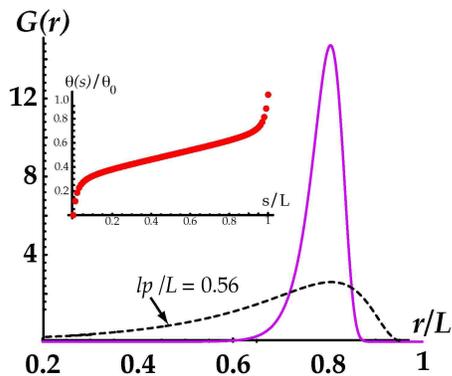} \caption{Comparison of the
radial distribution function of a PE (solid line) for $\kappa L=1$,
$\ell_{p0}/L=0.01$, and $\beta L=360$, with that of a neutral chain
(dashed line).  As the inset shows, when the two distributions do not
match, the equilibrium configuration of a bent PE is not given by a
constant-curvature profile.}
\label{fig:compare2}
\end{figure}

With the use of the radial distribution function, we have been
able to compare the statistical conformations of PE's with those
of uncharged \cite{frey} wormlike chains. Figure
\ref{fig:compare2}, displays the PE end-to-end distribution (solid
line) along with the WLC distribution (dashed line) in a case in
which it is not possible to collapse the two distributions on top
of each other.  The persistence length of the neutral WLC in the
figure was adjusted so that the location of the maxima of the two
distributions are the same. The plot of the uncharged WLC is for
$\ell_{p}/L=0.56$.  The distribution is for a PE segment with
$\ell_{p0}/L=0.01$, $\kappa L=1$, and $\beta L=360$.

Using these parameters, we also calculated the energy-minimizing shape
of a PE [Eq.  (\ref{gentheta})] as shown in the inset of Fig.
\ref{fig:compare2}.  It is obvious that end effects are not negligible
in this case and that the response of the PE to the bending force is
different from that of neutral chains.  This example indicates a
correlation between regimes in which the statistical conformations of
a PE chain and that of a WLC differ and circumstances under which the
classical, energy-minimizing shape of a PE segment does not trace out
the arc of a circle.

\begin{figure}[tbp]
\includegraphics[height=1.5in]{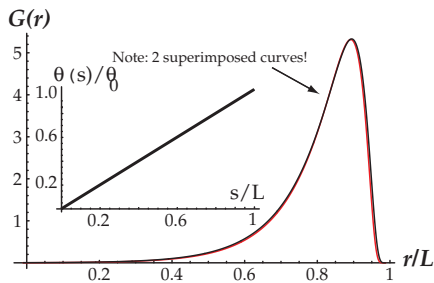} \caption{Comparison of the
radial distribution function of a PE for $\kappa L=100$,
$\ell_{p0}/L=0.0001$, and $\beta L=36000$, with that of a neutral
chain with the intrinsic persistence length $\ell_{p}/L=0.876$.  The
inset shows that when the two distributions collapse on top of each
other, the PE bends with constant curvature at equilibrium.}
\label{fig:compare1}
\end{figure}

There also exist regimes in which the conformational statistics of
PE chains in the rod-like limit are identical to those of WLC's
with adjusted persistence lengths \cite{en2}.  For such cases, PE
and WLC distributions are indistinguishable to the unaided eye.
Figure \ref{fig:compare1} illustrates an example of this regime.
The distribution function of the PE with $\ell_{p0}/L=0.0001$,
$\kappa L=100$ and $\beta L=36000$ completely obscures the
distribution function of a WLC with the intrinsic persistence
length $\ell_{p}/L=0.876$.  As in the previous example, the
energy-minimizing shape of the PE is shown in the inset. It is
clear that the energy-minimizing shape of the PE is not
distinguishable from that of a neutral chain, as the PE also bends
with a constant curvature in this example.

We have found that whenever there is a virtually perfect collapse
of the distribution function of a PE onto that of a neutral chain,
the persistence length of the neutral chain follows Odijk's
prediction, in that, $\ell_{p}=\ell_{e}+\ell_{p0}$, where
$\ell_{p}$ is the effective persistence length of the charged
chain, and $\ell_{e}=\ell_{\rm Odijk}$ \cite{odijk}. It is
noteworthy that in these regimes the energy-minimizing shape of
PE's is an arc of circle in accordance with the approximation
utilized by Odijk in his derivation of Eq. (\ref{eqodijk}).

Figure \ref{fig:newper} compares the values of electrostatic
persistence length, $\ell_{e}$ obtained by radial distribution
functions (hollow circles) to Odijk's formula (straight solid
line). In the figure, the electrostatic persistence length based
on Eq.  (\ref{eq:l01}) is also plotted (triangles).  For small values of
$\beta L$, $\ell_{e}$'s obtained through two different noted methods,
coincide with Odijk's persistence length, $\ell_{\rm Odijk}$.  As
$\beta L$ increases, the deviation of $\ell_{e}$ from $\ell_{\rm
Odijk}$ becomes more significant.  However, the electrostatic
persistence length obtained through the radial distribution function
and the one found by using the ``real'' shape of the chain match each
other quite well.

As we decrease the quantity $\kappa L$, the persistence lengths
obtained by distribution function and energy-minimizing shape also
start to deviate from each other.  This points to the fact that
replacing a PE chain with a WLC with an adjusted persistence length is
not well-justified in all regimes and that one should use care in the
utilization of the notion of an electrostatic persistence length.

\section{Conclusions}   \label{sec:conc}

Our investigation of the equilibrium shape of a bent PE has
yielded three striking results.  The first is the fact that the
Odijk formula, Eq. (\ref{eqodijk}), for the persistence length
applies almost perfectly to the case of a fluctuating PE when the
bent equilibrium segment has a constant curvature. This is
consistent with one of the fundamental assumptions underlying the
derivation by Odijk \cite{odijk}.

The second result is a suggestion for an improved calculation of the
persistence length based on the energy of the bent PE. This approach
appears to yield results in much closer accord with the calculations
of the radial distribution function of fluctuating PE segments, even
in regimes in which Eq.  (\ref{eqodijk}) does not work.  We find that
the fundamental tactic of extracting a persistence length from the
equilibrium energy of a bent PE yields excellent predictions for the
effective persistence length of an ensemble of fluctuating PE's over a
very wide range of parameters---if, however, one performs a
conscientious calculation of the actual shape of the bent PE.

Finally, we are able to characterize the shape of the distorted PE
segment in terms of ``shoulder'' regions, immediately adjacent to the
end-points of the chain, at which the curvature is significantly
greater than in the chain's interior.  It seems highly probable to us
that issues of PE energetics are intimately connected to the
quantitative features of these shoulder regions.  We are not yet able
to claim complete resolution of the questions associated with the
energetics and conformational statistics of rod-like PE chains.
However, the fact that one can, at least in principle, systematically
investigate the equilibrium properties of a torqued PE segment gives
rise to the expectation of substantial progress in the
characterization of the action of the important biomolecules in the
family of PE's.

The authors would like to acknowledge helpful discussions with W.M.
Gelbart, M. Kardar, R.R. Netz, I. Borukhov, B. Bozorgui, H. Diamant,
K. -K. Loh, V. Oganesyan, and G. Zocchi.  This research was supported
by the National Science Foundation under Grant No.  CHE99-88651.  One
of us (R.G.) would like to thank the group of Prof.  de Gennes at
Coll\`ege de France for its hospitality and support during his
visit.

\appendix

\section{Expression of the Energy of the bent PE in a Cosine Basis
Set} \label{Coulomb}

In this Appendix, we outline the method by which one
expands the energy of the bent PE in a basis set that automatically
satisfies the free boundary conditions at the end of the rod.   We
begin by expressing the distortion of the rod in terms of the
two-dimensional vector $\textbf{a} = (t_{x}, t_{y})$.  This means that
\begin{equation}
{\bf t}(s)=\frac{(a_{x}(s),a_{y}(s),1) }{
\sqrt{1+a_{x}^{2}(s)+a_{y}^{2}(s)}},\label{t-a}
\end{equation}
Using the Fourier representation of the screened Coulomb interaction
we find
\begin{widetext}
\begin{eqnarray}
\int_0^L d s \int_0^L d s' \frac{e^{-\kappa |{\bf r}(s) - {\bf
r}(s^{\prime})|}}{|{\bf r}(s) - {\bf r}(s^{\prime})|}&=&\int
\frac{d^3 {\bf k}}{(2 \pi)^3} \frac{4 \pi }{ k^2+\kappa^2} \int_0^L
d s \int_0^L d s' \; \exp\left\{i {\bf k} \cdot \left[ {\bf r}(s)
- {\bf r}(s^{\prime})\right]\right\} \nonumber \\
&\simeq&\int \frac{d^3 {\bf k}}{(2 \pi)^3} \frac{4 \pi}{
k^2+\kappa^2} \int_0^L d s \int_0^L d s' \; \exp\left\{i {\bf
k}_{\perp} \cdot \int_{s}^{s'} d u \; {\bf a}(u)-\frac{i k_z }{ 2}
\int_{s}^{s'} d u \; {\bf a}(u)^2+ i k_z (s'-s)\right\} \nonumber \\
&=&\int \frac{d^3 {\bf k}}{(2 \pi)^3} \frac{4 \pi }{ k^2+\kappa^2}
\int_0^L d s \int_0^L d s' \; e^{i k_z (s'-s)} \nonumber \\
&& \;\times\; \left[1-\frac{1 }{ 2 }\left({\bf k}_{\perp} \cdot
\int_{s}^{s'} d u \; {\bf a}(u)\right)^2-\frac{i k_z }{ 2}
\int_{s}^{s'} d u \; {\bf a}(u)^2+O(a^3)\right].
  \label{1distance}
\end{eqnarray}
\end{widetext}
The quantity $\textbf{k}_{\perp}$ is the projection of the wave vector
$\textbf{k}$ on the $x$-$y$ plane.  The quantity $k_{z}$ is the
$z$-component of that three-dimensional vector.  Next, we use the
series expansion ${\bf a}(s)=\sqrt{2}\sum_{n=0}^\infty {\bf A}_{n}
\cos \left(\frac{n \pi s}{L}\right)$ as appropriate for the open-end
boundary condition, and assume for simplicity that ${\bf r}$ is, on
average, oriented along the $z$-axis so that ${\bf A}_0=\frac{1 }{
\sqrt{2}}\int_0^L d s \;{\bf a}(s)=0$.  This leads to the following
representation of the Hamiltonian of the PE rod
\begin{equation}
 \frac{\cal E}{k_{\rm B} T}= \frac{\ell_{p0}}{2L}\sum_{n=1}^{\infty}(n
\pi)^{2} {\bf }A_{n}^{2}+\frac{\beta L}{2}\sum_{n,m=1}^{\infty}{\bf
A}_{n} \cdot {\bf A}_m E_{nm}, \label{dist7}
\end{equation}

\noindent where

\begin{widetext}

\begin{eqnarray}
E_{nm}&=&{4 L\over \pi n m}\int_{-\infty}^{\infty} {d k_z  \over 2
\pi} \left(k_z^2+\kappa^2\right)
\ln\left[\frac{(\pi/d)^2+\kappa^2+k_z^2}{\kappa^2+k_z^2}\right] \nonumber \\
&\times& \left\{\cos \left[{\pi \over 2}(n-m)\right]
\left[\frac{\sin\left({k_z L \over 2}\right)\sin\left({k_z L \over
2}-{\pi (n-m) \over 2 }\right)}{k_z \left[k_z-{(n-m) \pi/L
}\right]} -\frac{\sin\left({k_z L \over 2}-{\pi n \over 2}\right)
\sin\left({k_z L \over 2}-{\pi
m \over 2}\right)}{(k_z-{n \pi/L}) (k_z-{m \pi/L})}\right] \right. \nonumber \\
&&\left.-\cos \left[{\pi \over 2}(n+m)\right]
\left[\frac{\sin\left({k_z L \over 2}\right)\sin\left({k_z L \over
2}-{\pi (n+m) \over 2}\right)}{k_z [k_z-(n+m) \pi/L]}
-\frac{\sin\left({k_z L \over 2}+{\pi n \over 2}\right)
\sin\left({k_z L \over 2}-{\pi m \over 2}\right)}{(k_z+n \pi/L)
(k_z-m \pi/L)}\right]\right\},\label{Enm}
\end{eqnarray}

\end{widetext}
are the elements of the electrostatic energy matrix in the
cosine basis set. It is now sufficient to replace $a_{x}(s)$ by
$\theta_{x}(s)$, and similarly for $a_{y}(s)$.

A thorough investigation of the energy matrix (Eq.  (\ref{dist7})) is
given in Refs.  \cite{rudnick,roya}.  The requirement that the coarse
graining length $b$ not exceed the smallest wavelengths appearing in
the cosine basis set puts a restriction on the size of the matrix
energy.  If the length of the PE is $L$, this means that the size,
$N$, of the basis set satisfies $N \le L/b$.  At no point in our
calculations was this inequality violated.

An advantage of the cosine basis set, quite aside from automatic
satisfaction of the open boundary conditions, is that when $n$ and
$m$ are large, the matrix elements in Eq. (\ref{dist7}) are dominated
by those for which $m=n$. This reflects the dominance of elastic
energy at short wavelengths.

\section{Calculation of the Minimum Energy of a Bent Rod}
       \label{minenergy}
In this Appendix, the minimum energy of a charged chain that is
slightly deformed about the rodlike configuration with the use of
Eq. (\ref{invker}) is derived.  We begin with the expression for
the angle when the ends of the rod have been torqued:
\begin{equation}
\theta(s) \propto K(s,0)-K(s,L).
         \label{eq:theta1}
\end{equation}
Here the total arclength of the rod is assumed to be $L$.  The
relationship between the kernel $K(s,s^{\prime})$ as given by Eq.
(\ref{invker}) and the energy operator ${\cal H}(s,s^{\prime})$ as
given by Eq. (\ref{op1}) is
\begin{equation}
\int_{0}^{L}ds^{\prime \prime} \ {\cal H}(s,s^{\prime
\prime})K(s^{\prime \prime},s)=\delta(s-s^{\prime}).
         \label{eq:KandH}
\end{equation}
To obtain the proportionality constant in Eq. (\ref{eq:theta1}),
let us assume that the angle at $s=0$ is $-\theta_{0}/2$, while
the angle at $s=L$ is $\theta_{0}/2$. Then, we have
\begin{equation}
\theta(s) = \frac{\theta_{0}}{2} \
\frac{K(s,0)-K(s,L)}{K(0,L)-K(0,0)}.
         \label{eq:equality}
\end{equation}
The above kernel is symmetric, in that $K(x,y)=K(y,x)$;
furthermore, there is reflection symmetry in the looped rod in
that $K(0,0)=K(L,L)$.  The next step is to note that the energy of
the rod is the expectation value of the energy operator, i.e.
\begin{equation}
\frac{\cal E}{k_{\rm B} T} = \frac{1}{2} \int_{0}^{L} \int_{0}^{L}
ds ds^{\prime} \ \theta(s) {\cal H}(s,s^{\prime})
\theta(s^{\prime}).
         \label{eq:en2}
\end{equation}
If we plug in the solution (\ref{eq:equality}) for $\theta(s)$ in
Eq. (\ref{enop}) and
make use of the relation (\ref{eq:KandH}), we end up with the
expression in Eq. (\ref{eq:en3}) for the energy of the bent rod.

\bibliography{shape}

\begin{thebibliography}{22}
\bibitem[{en1()}]{en1}
\bibinfo{note}{Numerical evaluation of $G(r)$ based on a Monte-Carlo simulation
  in Ref. \cite{frey} reveals that the approximation taken for analytical
  calculation of $G(r)$ is good when $L/\ell_{p}\leq 2$. Thus we restrict our
  attention to this regime and assume that $\ell_{p0}/L= 0.5$ is the onset of
  rodlike limit.}
\bibitem[{en2()}]{en2}
\bibinfo{note}{In Refs. \cite{rudnick,roya}, we have identified the regimes in
  which the conformational statistcs of a PE differ from those of a WLC and
  those in which the PE behaves like a WLC.}
\end{thebibliography}

\end{document}